**Location-Based and Audience-Aware Storytelling:**
*Grace Plains* **and** *Bodies for a Global Brain*

Jeff Burke, Jared J. Stein
Center for Research in Engineering, Media and Performance (REMAP)
UCLA School of Theater, Film and Television (TFT)
{jburke, jared}@remap.ucla.edu

September 25, 2016

Everyday life now includes frequent exposure to media that changes depending on algorithms developed by content providers. While the daily user of digital, Internet-enabled devices has some explicit control over what they read and see, the providers fulfilling searches, offering options, and presenting material are using increasingly sophisticated real-time algorithms that tune and target content for the particular user. Shopping, research, news, entertainment, social media, and other services select, arrange and deliver content dynamically based on sets of variables—including users' past use, purchasing and viewing habits, current or recent locations, and even contact lists. (See, for example, Brusilovski, Kobsa, & Nejdl, 2007.) With shifting text and layout, embedded advertisements, sponsored content in news scrolls, menu choices, etc., the algorithms at play primarily serve providers' interests. They also redefine the historical relationships between tellers and users, providing a responsiveness paralleled only by forms of live performance incorporating elements of improvisation and audience interaction. The general accessibility of algorithmically driven content delivery techniques suggests significant untapped potential for new approaches to narrative beyond advertising and commercially orientated customization. At UCLA REMAP, faculty, staff, and students have been investigating the applications and implications of such constructs within the dramatic arts. (Burke et al, 2006; Burke, 2014)

Exploring these ideas in 2013–14, two full-scale experimental productions resulted from a two-quarter course created as part of UCLA TFT's participation in Google's Glass Creative Collective: 1] *Grace Plains* by Cole Baker, Pierre Finn, and Phill Powers, which combined participatory, immersive theater with live-action role-playing, viewed in-person and live online; and 2] *Bodies for a Global Brain* by Eben Portnoy, a web cinema piece that used machine learning to provide dialogue to its actors dynamically. The specific impetus for the experimentation was to explore Glass as a storytelling tool—given its point-of-view recording, wireless Internet access, and capability to deliver (algorithmically determined) digital information to individuals, received hands-free. The creation of these pieces involved faculty, students, research staff, and software developers. From the outset, they aimed to establish a range of creative possibilities for the use of Glass: location-based and audience-aware narrative conventions, methods of telling and experiencing a story that shift dynamically based on viewers' locations, identities, and choices. Ultimately, the development process yielded two different productions that were distinct in platform and format, but similar in theme and the challenges presented to creators, casts and crews.

*Bodies for a Global Brain* integrated a traditional episodic story structure with text that was dynamically selected from social media using machine learning, which was delivered to its primary performers via Glass. Shot in public spaces in Los Angeles, these two actors used the text as dialogue to interact with unknowing passersby, other actors not wearing Glass, and each other—as the writer-director manipulated the parameters of text selection and delivery. Text mined from social media served as the dialogue, re-contextualized by and motivating changes in the action of the scenes, in real time.

*Grace Plains*' in-person audiences were participants in a live-action role-playing game facilitated by Glass. They were led along a journey of continuously evolving circumstances at YouTube Space Los Angeles, across multiple studios and public spaces, which collectively became the project's set. Groups of eight people (six audience members and two actors) were fed, via Glass, suggested dialogue, motivations, facts, and options that progressed the story and overall experience.

The course, *Location-Based and Audience-Aware Storytelling*, taught by one of the article's authors (Jeff Burke) and supported by REMAP research staff, was designed to explore how to conceptualize, write, produce, and then distribute dramatic content that takes advantage of the capabilities of wearable technology (like Glass) to expand narrative possibilities. The resulting work, which included a series of smaller projects leading up to the two full-scale productions described here, melded "systems thinking" with "story thinking." Like the daily experience of interacting with contemporary media, but involving human performers rather than only web browsers or mobile apps, both *Bodies for a Global Brain* and *Grace Plains* strove to integrate dynamic variables into fairly conventional narratives—i.e., technologically-supported shifts in perspective or story progressions, especially those that take into account real-time information about the audiences and their locations. Of critical importance to the student authors became the possibility to deliver, privately, customized information to each individual wearing Glass while they were participating in or enacting a story within a shared space.

*Grace Plains* cast its in-person, Glass-wearing audiences as small groups of visitors to an artificial intelligence laboratory. For each performance, each of these audience members was given one of six individual character identities—senator, investor, documentarian, international thief, computer engineer, or scientist. Along the fictional VIP tour, they were introduced to the lab's government-appointed overseer and its chief scientist (the two actors also wearing Glass), who were generally at odds about the objectives of the lab's research. When a seemingly routine issue arose offstage, the visitors were left alone to interact, as their characters, with the lab's primary project, an artificially created mind voiced by a third actor. As the tour continued through this entity's lens and sentiments, a dead body was found—heightening the tension between the overseer and chief scientist, and posing the questions of who was to blame and, ultimately, whether or not the artificial intelligence should exist and, if so, for what purpose. Within the world of the story, resolving these issues became the primary objective of the audiences' characters.

The script prescribed that each performance include a few mandatory plot pots, assuring that certain types of actions progress the event towards one of three possible conclusions (for the artificial intelligence to be destroyed, remain an instrument of the government, or be set free). However, the actions in between, building towards these targeted points, varied performance by performance with numerous choices being made in real time. A team of writers, in a control room elsewhere in the complex, fed the audience and performers selections of preexisting text, as well as new text written on the fly. This control room received live POV video from each Glass device, as well as live video from surveillance cameras, so these writers could follow both individual and group activity. The actors and audience-participants improvised in reaction to what was happening in the playing spaces and what they were receiving from the writers via Glass.

Though much was prewritten and pre-rehearsed, the real-time reactions of the writers and actors were critical. When necessary to keep the story's general progression on track, the writers' room was able to collude with the actors, including the third actor voicing the "AI" from another control room, but even this was steered by choices made in reaction to the unique circumstances of a particular performance. What and when to send via Glass, and to whom, had to be constantly monitored—whether the content itself was pre-scripted or improvised. From the beginning—when the lab's visitors first received information about their characters' backstory—up until the final choice to kill, hold hostage, or liberate—the writers experimented with new revelations and clues, and suggestions for dialogue, character choices, and allegiances. In a third control room, a video-switching team selected from all available cameras, including Glass POVs, to generate a single live stream of the event, for YouTube Live audiences and the monitors in the public areas of YouTube Space Los Angeles.

*Bodies for a Global Brain* was crafted as the first two short episodes of a hypothetical, longer web series. In the not-so-distant future, a young programmer, Ada, and her new boyfriend, Ray, experiment by allowing their actions and interactions to be controlled by an evolving global intelligence. Having initially met on a social media site, for the first time they meet in person, they agree to wear Glass and relinquish their emotional autonomy to the will of a new zeitgeist, the Internet's total sum of accumulated thoughts. These characters venture out into the public, to share thoughts that are not their own, the phrases and short sentences fed to them. They must also use these thoughts to communicate what they themselves actually feel and need to express.

The project sought to explore how certain elements of a story's creation (in this case, its dialogue) could *actually* emerge from a "collective consciousness" of social media, and attempted to build a prototype system that, in effect, became something analogous to Ada's proposition within the story. It harvested text from Twitter, and organized these tweets according to specific emotional intents—using machine-learning algorithms[1] created by the class and trained according to rules developed by the writer-director. The actors playing Ada and Ray interacted with each other while limited in their dialogue to the current tweet being fed to each of them independently by the system. They also interacted with other actors not wearing Glass. These various scenarios were rehearsed—first with scripted dialogue, then with manually selected tweets, and then with automatically selected tweets. Other times, Ada and Ray interacted with the public. As the two young people wearing futuristic-looking goggles thrusted random fragments of thought at these unknowing recipients, they were met with varied responses, to which they could vocally respond using only more seemingly random fragments. In all cases, the text that Ada and Ray used for verbal communication was unknown—by them and everybody else in the cast and crew—until delivered by the machine-learning algorithm.

Rather than choosing the specific dialogue directly, the writer-director selected (live) one or more emotional intents—for example, "love and obey" or "deflect and defy"—which was used by the algorithm for tweet selection. Thus, what text was fed to the two actors, via Glass, was guided via human decisions about how to propel a scene's emotional trajectory, yet selected algorithmically.

---

[1] The specific approach applied sentiment analysis techniques (Pang & Lee, 2008), along with manual training.

*Bodies for a Global Brain* was performed live and recorded from a traditional third-person camera perspective, with the POV cameras of Glass providing additional coverage from the perspectives of Ada and Ray. For these pilot episodes, the control of text selection was kept on set with the writer-director—to sway the actors towards specific objectives or tactics, or to introduce, reverse, or exacerbate obstacles. The writer-director's eventual goal, however, was for online audiences to steer Ada and Ray's exploration of the world and interaction with each other, through a web interface that enables online participants to help adjust a scene's emotional intent or other variables, while leaving key trajectories of the scene intact.

Both projects' long-term aspirations included engaging online audiences in the customization of the very narratives they would be receiving—with the large-scale influence (described above) for *Bodies for a Global Brain* and as additional, remote role players for *Grace Plains*. In this way, both projects explored, in new forms, ongoing negotiations between content providers and content consumers in contemporary media.

Further, in both cases, the authors continued writing, or adjusting their writing, in real time to provide and respond to dynamically shifting variables. They operated "systems" in which their original scripts played parts, but expanded to embody the algorithmic relationships expressed in code. The code in these systems was not just making decisions "behind the scenes" based on each story's algorithms; they involved human-computer interfaces that expanded on and incorporated the initial scripts, making them centerpieces of digital systems used to control the progressions of the stories.

The primary interface that developed from the *Grace Plains* script was a carefully customized website for real-time text delivery never seen by the audience. It served as a control interface through which the writers could efficiently deliver both prewritten and improvised text to one or more characters (audience or actors). *Bodies for a Global Brain* used an on-set web interface, viewed alongside a traditional script, for selecting emotional intents and triggering text. It also used a traditional code-based interface to configure and train the supervised learning algorithm that inserted tweets into the action as dialogue, delivered to the performers via Glass.

Both pieces suggest an important, nascent role for comingling code and dramatic writing. They suggest how it may result not only in new story systems but also unique human-computer interfaces for authors, performers, and others—perhaps initially based on scripts with familiar, traditional forms (descriptions of settings and characters, actions and dialogue) and later incorporating aspects of computational narrative. (Mani, 2012) While *Grace Plains* and *Bodies for a Global Brain* were being written, the interfaces incorporating the scripts had to be conceived anew, uniquely designed for the stories being told. These *interfaces* were used to control and influence variables in real-time. Underneath remained *texts* (albeit dynamic ones) in the traditional sense, directing the dramatic action of human performers, while also—through code—directing the action of each piece's digital systems.

Today, we routinely encounter content that is selected, manipulated, and/or customized via algorithms. These algorithms and the content they deliver have increasing influence on almost every aspect of personal and private life. *Grace Plains* and *Bodies for a Global Brain* only scratch the surface of how storytellers and coders could collaborate to wield new approaches and the interfaces needed to do so. They do, however, imply important new research directions that combine the expressive capabilities of written dramatic action, millennia old, with the expressive capabilities of code, only just now being explored.

**Acknowledgements**

The projects *Grace Plains* and *Bodies for a Global Brain* were collaboratively conceived and created as attributed above, in collaboration with all of the students of the "Glass Class" taught by Jeff Burke at the UCLA School of Theater, Film and Television in 2013-2014: Cole Dalton Baker, Karan Chugh, Joon-Sub Chung, Pierre Finn, Ning Ji, Edi Leung, Eben Portnoy, Phillip Powers, Jeff Rosenberg, Zoe Sandoval, Tyler Schwartz, XiaoWei Smile, and Jackie Watson—with critical support from REMAP research staff Randy Illum, Alex Horn, Emily Holden, Jared J. Stein, and Zhehao Wang.  This work was supported in large part by Google through the Glass Creative Collective and a Faculty Research Award.